\documentclass{article}
\usepackage[english]{babel}
\usepackage{amsfonts,amssymb, amsmath}

\begin{document}

\title{On integrable by Euler planar differential systems}
\author{A.\,V.~Tsiganov\\
\it\small St. Petersburg State University, St. Petersburg, Russia;\\
\it\small Beijing Institute of Mathematical Sciences and Applications, Beijing, China;\\
\it\small e--mail: andrey.tsiganov@gmail.com
}
\date{}
\maketitle

\begin{abstract}
The subject of our discussion is the theory of differential equations as set out in two classical Euler's textbooks "Institutiones Calculi Differentialis" and "Institutiones Calculi Integralis". 
\end{abstract}

\section{Introduction}
\setcounter{equation}{0}
There is a vast literature regarding the computation of first integrals of planar differential systems
\begin{equation}\label{p-vf}
\frac{dx}{dt} = A(x, y)\qquad\mbox{and}\qquad  \frac{dy}{dt} = B(x, y)\,.
\end{equation}
By definition first integral $f(x,y)$ of equations (\ref{p-vf}) satisfies equation
\begin{equation}\label{int-eq}
X(f)=A(x,y)\partial_x f+B(x,y)\partial_y f= 0\,,
\end{equation}
where $X$ is the planar vector field
\[
X=A(x,y)\partial_x+B(x,y)\partial_y\,.
\]
Sadly, almost all modern works fail to mention Euler's integrability condition of equations (\ref{p-vf}), which became the starting point for further research of complete Pfaffian equations due by Jacobi, Liouville, Clebsch, Lie, Frobenius, Darboux and Cartan. 

Instead of (\ref{p-vf}) Euler studied equivalent differential equation 
\begin{equation}\label{eq-pf}
\omega=P(x, y)dx+Q(x, y)dy=0\,,
\end{equation}
replacing vector field $X$ on the one-form $\omega$ at $P=B$ and $Q=-A$, see Euler's papers \cite{eul} and textbooks \cite{eul-diff, eul-int}. Here and throughout, we use Euler's notations.

According to Theorem \S459 \cite{eul-int}, it only makes sense to speak of the integration of equation $\omega = 0$ (\ref{eq-pf}) when differential 1-form $L\omega$ is exact for some nonvanishing factor $L = L(x,y)$, i.e. 
\begin{equation}\label{eul-eq}
L\omega  = d\varphi(x,y)\,.
\end{equation}
or
\begin{equation}\label{m-eq}
LP=\frac{\partial \varphi}{\partial x} \qquad\mbox{and}\qquad LQ=\frac{\partial \varphi}{\partial y}\,.
\end{equation}
By differentiating (\ref{m-eq}) by $x$ and $y$ and then subtracting, Euler obtained   
\begin{equation}\label{eul-main}
\frac{\partial}{\partial y}\,LP= \frac{\partial}{\partial x}\, LQ\,.
\end{equation}
By multiplying (\ref{m-eq}) on $Q$ and $P$ and then subtracting, Euler obtained equation (\ref{int-eq}) 
\[
-Q\partial_x \varphi + P\partial_y \varphi=
A(x,y)\partial_x \varphi +B(x,y)\partial_y \varphi= 0\,.
\]
In \cite{eul,eul-diff,eul-int}, Euler discusses many various examples of integrable differential equations $\omega=0$ with polynomial, rational, algebraic and transcendental first integrals.  

Bear in mind that, at this time, there is no sensitivity to the distinction between local and global results, and that the actual mathematics is being performed at a strictly local level. This means that the function $f(x,y)$ can be multi-valued in $x$ and $y$, generating single-valued differential equations with a single-valued complete solution. So, in classical textbooks \cite{jac,lie-book,lie-book2,car}, authors prefer to discuss the existence of complete solutions (complete integrability), the existence of an infinitesimal transformation group (non-commutative integrability) and the existence of absolute or relative integral invariants instead of the existence of first integrals. 

In the context of modern topological mechanics, multi-valued functionals and non-exact closed differential 1-forms are discussed in \cite{nov82}.  The notion of a multi-valued functional originates from the topological study of the quantisation process of the motion of a charged particle in a Dirac monopole field, the Aharonov-Bohm experiment, Kirchhoff-type systems, etc.

\section{Few classical Euler's examples}
Let us briefly discuss a few of Euler's examples from Chapters II and III of the textbook \cite{eul-int}, partial cases of which have recently been reopened by modern authors. Naturally, it is not possible to consider all of Euler's examples in this note, including his famous integral for elliptic functions from Chapter VI of the same textbook. 

\subsection{Homogeneous equations}
In \S477 Euler considers equation $\omega=0$ when  $P(x, y)$ and $Q(x, y)$ are homogeneous functions of degree $n$
\[
P(\alpha x,\alpha y)=\alpha^n P(x,y)\qquad\mbox{and}\qquad
Q(\alpha x,\alpha y)=\alpha^n Q(x,y)\,,
\]
which satisfy to equations 
\begin{equation*}  
x\partial_x Q(x, y) +y\partial_yQ(x, y)=n Q(x,y)\,,\qquad
x\partial_x P(x, y) +y\partial_yP(x, y)=n P(x,y)\,.
\end{equation*}
according to Euler's theorem \cite{eul-int}. For $n\neq-1$, Euler substitutes $P=x^nU(u)$ and $Q=x^nV(u)$ into the 1-form
\[\omega=Pdx+Qdy=x^n(U+Vu)dx+x^{n+1}Vdu\]
where $u=x/y$ and $dy=udx+xdu$. 

He then observes that dividing this form by $x^{n+1}(U+Vu)$ yields an exact differential. 
So,  in this case multiplier is equal to 
\[
L(x,y)=\frac{1}{x P+yQ}\neq 0\,,
\]
whereas first integral $\varphi(f)$ of homogeneous planar differential system is defined by  
\[
f(x,y)=
\int  \frac{P}{xP  +  yQ}dx+\int \frac{Q}{x P+yQ} dy+
\iint
\frac{(n+1)QP - xP\partial_xQ-yQ\partial_y P }{( xP  +  yQ)^2}dxdy\,.
\]
Function $\varphi=\exp(\lambda f)$ is the homogeneous first integral of degree $\lambda$, see comments in the Russian translation of \cite{eul-int}. Discussion of the case $n=-1$ may be also found in \cite{eul-int}. 

In fact, here Euler uses separation of variables in the polar coordinates on a plane.  In similar way Euler also studied dynamics separable in parabolic and elliptic coordinates on a plane, as an example see Euler's works on the two-center problem in \cite{eul}.

In \S431 and \S465 Euler considers the following differential equation 
\[
\alpha y \,dx+\beta x\, dy=0
\]
when $P$ and $Q$ are homogeneous functions of degree $n=1$. In this case multiplier is equal to
 \[L=\frac{\varphi(x^\alpha y^\beta)}{xy}\,,\]
 where $\varphi(x^\alpha y^\beta)$ is the first integral. Euler later uses this first integral and multiplier to consider quasi-homogeneous differential systems.
 
In \S467 we can find equation
\[
dy+yXdx=0
\] 
where $X$ is a function on $x$. Its multiplier $L$ and first integral $\varphi(f)$ are 
\[
L=\frac{1}{y}\varphi(f)\,,\qquad\mbox{where}\qquad f(x,y)=e^{\int Xdx}y\,.
\] 
Modern classification of such first integrals involving undetermined integrals on arbitrary functions is discussed in \cite{neto3,com23}.

\subsection{Compound differential equations}
In \S464 Euler considers compound differential equation
\[
\omega=\omega_1+\omega_2=(Pdx+Qdy)+(Rdx+Sdy)=0
\]
with known multipliers for both $\omega_1=(Pdx+Qdy)$ and $\omega_2=(Rdx+Sdy)$ and proposes some construction of the multiplier and firs integral of equation $\omega=0$. 

As an example, in \S471-\S472 Euler studies the following equation
\begin{equation}\label{ex-1}
dy+Xydx=\mathfrak{X}y^ndx\,,
\end{equation} 
where $X$ and $\mathfrak{X}$ are functions on $x$. In this case first integral $\varphi(f)$ is a function on 
\begin{equation}\label{ex-11}
f(x,y)=\frac{y^{1-n}}{1-n}\,e^{(1-n)\int Xdx}-\int e^{(1-n)\int Xdx} \mathfrak{X}\,dx\,.
\end{equation}
Of course, the Euler's theory is completely incorporated into modern computer algebra systems. As an example, let us consider
equivalent to (\ref{ex-1}) differential system
\[
\dot{x}=-F(x)\qquad\mbox{and}\qquad \dot{y}=cy^n+y G(x)\,.
\]
Using the computer algebra systems \texttt{Maple} or \texttt{Mathematica} and a standard laptop, we can compute solutions to Euler's equation (\ref{eul-eq}) in 2-3 seconds. In this case $f$ is equal to (\ref{ex-11})
\[
f(x,y)=\frac{y^{1 - n}}{1-n} e^{(1-n) \int{\frac{G(x)}{F(x)}dx}}-c\int{\frac{e^{(1-n)\int\frac{G(x)}{F(x)}dx}}{F(x)}dx } 
\]
whereas multiplier has the form 
\[
L=\frac{(1 - n)e^{(1-n)\int \frac{G(x)}{F(x)} dx}}{y^nF(x)}\,\varphi(f)
\]
If $G(x)=x(x-1)$ and $F(x)=ax+b$ we obtain a hypergeometric function instead of these indefinite integrals. 
This classical result of Euler's was revisited in \cite{per25}.  Similarly, multipliers and first integrals of planar systems from \cite{per25a} can easily be obtained using computer algebra systems.

In \S431 and \S465 Euler considered the following quasi-homogeneous equation
\[
(Pdx+Qdy)+(Rdx+Sdy)=(\alpha y \,dx+\beta x\, dy)+x^my^n\left(\gamma y\, dx+\delta x\, dy\right)=0\,, 
\] 
where $P,Q$ and $R,S$ are homogeneous polynomials of degrees $n_{pq}=1$ and $n_{rs}=m+n-1$.
In this case multipliers have the following form  
\[
L_{pq}=\frac{\varphi_1(x^\alpha y^\beta)}{xy}\qquad \mbox{and}\qquad
L_{rs}=\frac{\varphi_2(x^\gamma y^\delta)}{x^{m+1}y^{n+1}}\,.
\]
If we take functions $\varphi_1$ and $\varphi_2$ so that $L_{pq}=L_{rs}$ we obtain a common multiplier and first integral
\[
\varphi\left(\frac{1}{\mu}x^{\mu\alpha}y^{\mu\beta}-\frac{1}{\nu}x^{\nu\gamma}y^{\nu\delta}\right)=\mbox{const}.
\]
where
\[
\mu=\frac{\gamma m-\delta n}{\alpha\delta-\beta\gamma}\qquad\mbox{and}\qquad
\nu=\frac{\alpha n-\beta m}{\alpha\delta-\beta\gamma}\,,
\]
see details in \cite{eul-int}. Modern symbolic computation systems are not yet able to reproduce this first integral without making additional assumptions about the values of the constants $\alpha,\beta,\gamma,\delta$ and $m,n$.

To finish discussion on integrable differential forms, Euler wrote in \S516 :\\
\textit{
"Haec fusius non prosequor, quia ista exempla eum in finem potissimum attuli, ut methodus supra tradita aequationes differentiales tractandi exerceretur; in his enim exemplis casus non parum difficiles se obtulerunt, quos ita per partes resolvere licuit, ut pro singulis multiplicatores idonei quaererentur ex iisque multiplicator communis definiretur; nunc igitur alia aequationum genera, quae per multiplicatores integrabiles reddi queant, investigemus."}
\vskip0.1truecm
Using AI translator we immediately get\\
\textit{
I will not pursue this further, for I have brought these examples primarily to the end that the method for handling differential equations, as set forth above, might be exercised; for in these examples, cases of no small difficulty presented themselves, which it was possible to resolve in parts in such a way that for each, suitable multipliers were sought and from them a common multiplier was defined; Let us now, therefore, investigate other types of equations that can be rendered by integrable multipliers." }\\
These other Euler'types of equations are discussed in textbook \cite{eul-int} and in various papers devoted to different  physical applications \cite{eul}. 

Let us now consider a modern example from \cite{neto1,neto2} when
\[P=3(2\alpha + 1)(x^2 - y^2) - 12xy\,,\qquad Q=-y^2 + 9x^2 + 4(2\alpha + 1)xy\]
and
\[
R=6y\,,\qquad S=-(4\alpha + 2)y - 4x\,.
\]
It is easy to solve Euler's equation (\ref{eul-main}) using  a computer algebra system and to find multipliers depending on  two functions $\varphi_{1,2}(f_{1,2})$ on first integrals $f_{1,2}(x,y)$
\[
L_{pq}=\frac{(2\alpha x + x - y)^2}{(3x^2 + y^2)^3}\varphi_1(f_1)\,,\qquad f_1=
\frac{y^3-(2\alpha + 1)^3x^3 + 3xy(2\alpha + 1)(2\alpha x + x - y)}{9x^4 + 6x^2y^2 + y^4}
\]
and
\[
L_{rs}=\frac{1}{y^{5/3}}\varphi_2(f_2)\,,\qquad f_2=\frac{x-(2\alpha + 1)y}{y^{2/3}}\,.
\]
By selecting functions $\varphi_1$ and $\varphi_2$ in such a way that $L_{pq}=L_{rs}$ we can try to obtain common factor and first integral of quasi-homogeneous polynomial equation
\[
\omega=\omega_1+\omega_2=(Pdx+Qdy)+(Rdx+Sdy)=0\,.
\]
Of course, in similar way we can divide a given equation on the three parts
\[
\omega=\omega_1+\omega_2+\omega_3
\] 
with known multipliers and so on. These modern "telescopes" are discussed in \cite{neto3,com23}.

Other algorithms to integration of quasi-homogeneous polynomial systems are discussed in \cite{ll25}, see also references within.

\subsection{Construction of polynomial equations with polynomial integrals}
In Chapter 3 of textbook \cite{eul-int} Euler proposed to construct integrable equations starting with a smooth function $\varphi(x,y)$ which yields a complete equation
\[
\tilde{\omega}=d\varphi=0\,.
\]
After that we can change form of this equation using a function $L (x,y)\neq 0$
\[\omega=L^{-1}\tilde{\omega}=0\,.\]
The most important Euler's examples may be found in \cite{eul-int}, while a complete list of the obtained equations can be found in \cite{eul}.

Let us consider quadratic differential equations 
\[\omega=P(x,y)dx+Q(x,y)dy=0\,,\] 
where $P(x,y)$ and $Q(x,y)$ are polynomials of second order in $x$ and $y$ with real coefficients
in the framework of the Euler theory. Modern constructions of such systems with polynomial first integrals are discussed in \cite{ts01} and \cite{ll09}.

Suppose that our aim is to construct a system of differential equations 
\[
\left\{
\begin{array}{c}
\dot{x}_1=a_1x^2+a_2xy+a_3y^2 
\\
\dot{x}_2=b_1x^2+b_2xy+b_3y^2
 \end{array}
 \right.
\]
having fifth order polynomial integral
\begin{equation}\label{planar-f5}
f(x,y)=c_5x^5 + c_4x^4y + c_3x^3y^2 + c_2 x^2y^3 + c_1xy^4 + c_0y^5\,.
\end{equation}
Here $a_i,b_i$ and $c_i$ are some undetermined coefficients of homogeneous polynomials $P(x,y)$, $Q(x,y)$ and $f(x,y)$.

Substituting polynomial $f(x,y)$ (\ref{planar-f5}) into the standard definition of first integral (\ref{int-eq})
\[
X(f)=(a_1x^2+a_2xy+a_3y^2)\frac{\partial}{\partial x}\, f(x,y)+(b_1x^2+b_2xy+b_3y^2)\frac{\partial}{\partial y}\, f(x,y)=0
\]
we obtain 7 algebraic equations of second order
\begin{align}
5a_1c_5 + b_1c_4=0\,,\quad a_3c_1 + 5b_3c_0=0\,,\quad a_1c_1 + 2a_2c_2 + 3a_3c_3 + 5b_1c_0 + 4b_2c_1 + 3b_3c_2=0\,,\nonumber\\
a_2c_1 + 2a_3c_2 + 5b_2c_0 + 4b_3c_1=0\,,\qquad
2a_1c_2 + 3a_2c_3 + 4a_3c_4 + 4b_1c_1 + 3b_2c_2 + 2b_3c_3=0\,,
\label{25-eq2}
\\
4a_1c_4 + 5a_2c_5 + 2b_1c_3 + b_2c_4=0\,,\qquad
3a_1c_3 + 4a_2c_4 + 5a_3c_5 + 3b_1c_2 + 2b_2c_3 + b_3c_4=0\,.
\nonumber
\end{align}
on 11 undetermined coefficients $a_1, a_2, a_3$, $b_1, b_2, b_3$ and $c_0,c_1,c_2,c_3$. 

Let us now substitute the same polynomials into the Euler integrability condition (\ref{m-eq})
\[
LP=\frac{\partial \varphi}{\partial x}\qquad\mbox{and}\qquad LQ=-\frac{\partial \varphi}{\partial y}\,.
\]
where
\[
L=m_1x^2+m_2xy+m_3y^2\,.
\]
These equations yield 10 linear equations on the same 11 undetermined coefficients, which we also present explicitly for sake of completeness
\begin{align}
a_1m_1 - c_4=0\,,\quad a_3m_3 - 5c_0=0\,,\quad
a_1m_2 + a_2m_1 - 2c_3=0\,,\quad a_2m_3 + a_3m_2 - 4c_1=0\,,\nonumber\\
b_1m_1 + 5c_5=0\,,\quad b_3m_3 + c_1=0\,,
\quad b_1m_2 + b_2m_1 + 4c_4=0\,,\quad b_2m_3 + b_3m_2 + 2c_2=0\,,\label{25-eqlin}\\
a_1m_3 + a_2m_2 + a_3m_1 - 3c_2=0\,,\quad
b_1m_3 + b_2 m_2 + b_3 m_1 + 3c_3=0\,.\nonumber
\end{align}
It is natural that we prefer to solve linear equations (\ref{25-eqlin}) depending on parameters $m_k$ since they require significantly fewer computational resources than solution of quadratic equations (\ref{25-eq2}) .

Solution of (\ref{25-eqlin}) is parameterized by $a_3$, $b_3$ and $m_{1,2,3}$ and desired differential system is 
\[
\left\{
\begin{array}{l}
\dot{x}_1=m_3(a_3m_1m_3 - 4a_3m_2^2 - 10b_3m_2m_3)x^2 - 5m_3^2(a_3m_2 + 4b_3m_3)xy + 5a_3m_3^3y^2\\
\\
\dot{x}_2= 5m_1m_3(2a_3m_2 + 5b_3m_3)x^2 - m_3(4a_3m_1m_3 - 6a_3m_2^2 - 15b_3m_2m_3)xy+5b_3m_3^3y^2 
\end{array}
 \right.
\]
whereas first integral is equal to
\begin{equation} \label{f2}
f=L^2\Bigl(\left(2a_3m_2+5b_3m_3\right)x - a_3m_3y\Bigr)\,.
\end{equation}
Of course, we can speed up calculations using continuous and discrete isometries of a plane.

In similar way we can construct differential systems with the homogeneous cubic right-hand sides
\[
\dot{x}=\sum_{k=0}^3 a_kx^ky^{3-k}\,,\qquad \dot{x}=\sum_{k=0}^3b_kx^ky^{3-k}
\]
having fifth order first integral $f(x,y)$ (\ref{planar-f5}). In this case, the multiplier $L(x)$ is a linear polynomial that can be reduced using rotation and translation to the following form
\[L=x\,.\]
Solving the corresponding linear system of algebraic equations, we obtain integrable differential equations
 \[
\dot{x}_1=x(c_4x^2 + 2c_3xyc_3 + 3c_2y^2)\,,\qquad
\dot{x}_2=-5c_5x^3 - 4c_4x^2y - 3c_3xy^2 - 2c_2y^3 
 \]
 and first integral
 \[
 f=x^2(c_5x^3 + c_4x^2y + c_3xy^2 + c_2y^3)\,.
 \]
  In summary, when we substitute the inhomogeneous quadratic polynomials $P(x,y)$ and $Q(x,y)$, the $N$th-order polynomial $\varphi(x,y)$ and a suitable rational function $L(x,y)$ into the Euler integrability condition (\ref{m-eq}), we obtain a system of algebraic equations with undetermined coefficients that can be solved using a computer algebra system. It is straightforward to generate a list of such integrable equations and train an AI to recognize some of their canonical forms.

\section{Conclusion}
In \cite{jac}, Jacobi proposed a direct generalisation of Euler's construction of complete differential systems, which we will briefly discuss using the original Jacobi notation.

By substituting any independent functions $f_1, f_2, ..., f_{n-1}$ on $n$ variables 
$x_1,\ldots,x_n$ and some nonvanishing multiplier $M(x_1,\ldots,x_n)\neq 0$ into the Jacobi equation (1) on the page 331 
in his {\OE}uvres compl\'{e}tes  \cite{jac}  
\begin{equation}\label{x-jac}
MX_k=A_k\,,\qquad \mbox{where}\qquad A_k =\frac{\partial(x_k,f_1,\ldots,f_{n-1})}{\partial( x_1,\ldots, x_n)}\,,
\end{equation}
we obtain partial differential equation having a complete solution
\begin{equation}\label{j-pde}
X_1(x_1,\ldots,x_n)\frac{\partial f}{\partial x_1}+\cdots+X_n(x_1,\ldots,x_n)\frac{\partial f}{\partial x_n}=0
\end{equation}
and the corresponding complete system of ordinary differential equations
\[
\frac{dx_1}{X_1}=\frac{dx_2}{X_2}=\cdots=\frac{dx_n}{X_n}\,.
\]
Then, using properties of functional determinants, Jacobi proved equation (3) 
\begin{equation}\label{m-jac}
\frac{\partial MX_1}{\partial x_1}+\cdots+ \frac{\partial MX_n}{\partial x_n}=0\,.
\end{equation}
Of course, at $n=2$ equations (\ref{x-jac}),  (\ref{j-pde}) and (\ref{m-jac}) coincide
coincide with the Euler equations (\ref{eul-eq}-\ref{m-eq}),  (\ref{int-eq}) and (\ref{eul-main}), respectively. 

In the classical textbooks by Lie \cite{lie-book, lie-book2} and Cartan \cite{car} this Jacobi's construction 
of complete differential equations using functional determinants was rewritten using infinitesimal transformation group and  non-exact differential 1-forms instead of functions $f_k(x_1,\ldots,x_n)$ which can be multi-valued functions on variables $x_1,\ldots,x_n$ as in the Euler examples. Much later, this classical Jacobi construction of complete differential systems was reproduced by Nambu \cite{nam73} in the framework of the Hamiltonian mechanics.

\vskip0.2truecm
The research was carried out with the financial support of the Ministry of Science and Higher Education of the Russian Federation in the framework of a scientific project under agreement No 075-15-2025-013 by St. Petersburg State University as part of the national project "Science and Universities" in 2025.
\vskip0.2truecm

\textbf{Conflict of Interests:} The authors declare that they have no conflicts of interest.

\end{document}